\documentclass[pra, reprint, showpacs, nofootinbib]{revtex4-2}
\usepackage{graphicx}
\usepackage{amsthm, amsmath, amssymb}
\usepackage[fonts]{faust}
\usepackage{dirac}

\usepackage[colorlinks=true,linkcolor=blue]{hyperref}
\usepackage{cleveref}

\newcommand\defn[1]{{\bfseries\slshape #1\/}}
\newcommand\hc{{\mathrm{hc}}}
\newcommand\ip{{\mathrm{ip}}}
\newcommand\mc{{\mathrm{mc}}}
\newcommand\pa{{\mathrm{pa}}}

\begin{document}

\title{The paraxial approximation in quantum optics I:\\ Henochromatic modes of a scalar field}

\date{January 15, 2023}

\author{Christopher Beetle}
\affiliation{Department of Physics, Florida Atlantic University, 777 Glades Road, Boca Raton, FL 33431}
\author{M.\ Fernanda \surname{Jongewaard de Boer}}
\affiliation{Department of Physics, Florida Atlantic University, 777 Glades Road, Boca Raton, FL 33431}

\begin{abstract}
This paper examines how best to associate quantum states of a single particle to different modes of a narrowly collimated beam of classical radiation modeled in the paraxial approximation.  Our analysis stresses the importance of the relationship between two inner products naturally arising in the problem.  These are the inner product used to expand a general beam as a superposition of orthogonal modes in the paraxial approximation, on the one hand, and the canonical inner product on which the statistical interpretation of quantum (field) theory is founded, on the other.  While several candidates for the sort of association between beam modes and single-particle quantum states have been proposed in the literature, here we argue that one of them is uniquely well suited to the task.  Specifically, the mapping of beam modes to the ``henochromatic'' quantum states previously introduced by Sudarshan, Simon and Mukunda \cite{SSMs, SSMv} is unique within a large class of similar mappings in that it is unitary in a mathematically precise sense.
\end{abstract}

\pacs{%
    03.70.+k,     % Quantized fields
    42.50.-p,     % Quantum optics
    42.50.Tx,     % Optical angular momentum (quantum optics)
    42.60.Jf      % Laser radiation characteristics 
}
\maketitle

The paraxial approximation models beams of classical radiation that are both narrowly collimated and nearly monochromatic, such as those produced by a laser.  Radiation fields in the paraxial approximation can be expanded as superpositions of various bases of orthogonal modes, such as the Hermite-- or Laguerre--Gauss modes, which correspond physically to different geometric shapes of the wavefront within the beam that are (relatively) easy to distinguish experimentally.  Such mode expansions suggest an intriguing opportunity at the quantum level.  Namely, the Hilbert space of possible shapes of a single photon's wavefront is infinite-dimensional, whereas the Hilbert space of its polarization states is only two-dimensional.  Thus, if one can ascribe the properties of beams in the paraxial approximation to single-particle particle states of a quantum field, then in principle one can encode a much larger set of quantum information in a single photon than one could using polarization alone.

The difficulty in a straightforward application of the program sketched above is that the paraxial approximation is, after all, an \emph{approximation}.  The fields it constructs on spacetime do not solve the wave equation $\Box\Psi(\vec r, t) = 0$ exactly.  In contrast, the field operator $\hat\psi(\vec r, t)$ in a free, scalar field theory does solve the wave equation exactly \emph{by definition}.  Clearly one cannot expand such an operator as a superposition (with operator-valued coefficients) of (c-number) modes that fail to solve the field equation, as one routinely does using the plane-wave basis of single-particle momentum eigenstates.  Thus, in order to realize the potential of exploiting modes of the wavefront shape, as understood in the paraxial approximation, to broaden the quantum information channel per photon, one must find a way of associating to each paraxial field mode a proper single-particle state (\textit{i.e.}, an exact, positive-frequency solution of the wave equation) of the quantum field.

Several proposals of such associations have appeared in the literature \cite{DG, AW, CPB, SSMs, SSMv, ACM, AML, AVNW}.  In the present paper, we argue that one of them, based on the ``henochromatic'' states introduced in \cite{SSMs, SSMv}, is uniquely well suited to the task.  In detail, we show that the linear mapping from orthogonal modes of a paraxial beam to henochromatic single-particle states is \emph{unitary}.  Physically, this implies that the quantum logic of the preparation and subsequent measurement of such quantum states exactly mirrors the familiar mode expansions of the paraxial approximation.  We also show that the resulting set of quantum states is (over)complete, so that every single-particle state can be expanded (uniquely) as a superposition of henochromatic states derived from any given basis of the space of solutions in the paraxial approximation.  Finally, we argue that the same optical elements used to isolate particular modes of a paraxial beam (\textit{e.g.}, forked diffraction gratings, spiral wave plates, \textit{etc.}) work essentially without modification to do the same for appropriate pulses constructed from henochromatic single-particle states.

This paper is organized as follows.  Section 2 reviews the paraxial approximation for scalar fields.  (A companion paper \cite{Vectorpaper} extends the present results to Maxwell fields.)  Section 3 recalls a few essential facts about the standard (Fock) quantization of a free, massless, scalar field .  Section 4 states and proves our key result, that the mapping from solutions of the paraxial wave equation to \emph{henochromatic}, single-particle, quantum states is \textit{unique} within a large class of similar mappings in that it preserves the inner product used to form the orthogonal mode expansions of a paraxial beam mentioned above.  Section 5 shows that the henochromatic fields associated to an orthogonal basis of beam modes are not only orthogonal, but also form a complete basis.  Section 6 summarizes our results and discusses the sense in which the geometry of the wavefront for henochromatic single-particle modes on spacetime resembles that of the underlying modes of a beam in the paraxial approximation.

\section{Paraxial approximation for classical scalar fields}\label{S:CPA}

A general radiation field in scalar field theory solves the homogeneous wave equation 
\begin{equation}\label{wEqIn}
 	 \Box \Psi ({ \bf r }, t) 
 	    := \biggl( \pdby[2]{x} + \pdby[2]{y} + \pdby[2]{z} - \frac{1}{c^2} \pdby[2]{t} \biggr) \Psi(\vec r, t) 
 	    = 0.
\end{equation}
The restriction to paraxial beams is most transparent if one writes (the positive frequency part of) the general solution of \cref{wEqIn} in its Fourier representation \cite{MB, EW}
\begin{align}\label{pwExp}
	\Psi^+(\vec r , t) 
        &= \int \ed^3 k\, \rho(\vec k)\, \mathcal{A}(\vec k)\, 
            \frac{\ee^{\ii (\vec k \cdot \vec r - c \|\vec k\| t)}}{\sqrt{(2 \pi)^3\, 2 \|\vec k\|\, \rho(\vec k)}}, 
\end{align}
where the scaling factor $\rho(\vec k) > 0$ built into the definition of $\mathcal{A}(\vec k)$ in this way describes the density of states in Fourier 3-space\relax 
%%%
\footnote{The density of states may be chosen freely by adjusting the normalizations of the plane-wave states superposed in \cref{pwExp} as shown.  The two most common choices are $\rho(\vec k) = 1$, corresponding to a uniform distribution of states in the Fourier 3-space of the particular inertial frame subsumed in \cref{pwExp}, and $\rho(\vec k) = 1 / 2\|\vec k\|$, corresponding to a uniform distribution of states on the positive-frequency light cone in Fourier 4-space.  The former leads to simpler commutation relations in the quantum theory and the latter, to better relativistic covariance properties.  The choice is largely irrelevant at the classical level, but it will be helpful to retain the freedom to consider different possibilities below.\relax
\label{fn:Dens}}\relax
%%% 
.  The paraxial approximation applies when the support of $\mathcal{A}(\vec k)$ is mostly confined to a small region around a given $\vec k_0 \ne \vec 0$ in Fourier space.  More precisely, this means that $\mathcal{A}(\vec k) \approx 0$ unless $\|\vec k - \vec k_0\| \ll k_0 := \|\vec k_0\|$.  We refer to the set of solutions of \cref{wEqIn} satisfying this loose criterion as the \defn{paraxial regime} of scalar field theory.

Fields in the paraxial \emph{regime} \cite{AW, AVNW, AML, CPB, ACM, SSMs, SSMv} solve \cref{wEqIn} exactly, and are approximately monochromatic by virtue of their localization in Fourier space.  In contrast, the paraxial \emph{approximation} models fields in the paraxial regime using a distinct set of fields \cite{DG} that are \emph{exactly} monochromatic, but solve \cref{wEqIn} only approximately.  The approximation can be derived by first restricting \cref{pwExp} to be exactly monochromatic, \textit{i.e.}, by choosing $\mathcal{A}(\vec k)$ to have (distributional) support on the sphere of radius $k_0$ about the origin in Fourier space.  Aligning the $+z$-axis with the principal wave vector $\vec k_0$, and using the transverse wave vector $\vec q := \vec k - \vec{\hat z \hat z} \cdot \vec k$ as coordinates on the (forward hemi)sphere, yields the strictly monochromatic field \cite{ACM}
\begin{equation}\label{mcMode}
    \Psi^\mc(\vec s, z, t) 
        = \int \frac{\ed^2 q}{2 \pi}\, \mathcal{F}(\vec q)\, 
            \ee^{\ii \vec q \cdot \vec s}\, 
            \ee^{\ii z \sqrt{k_0^2 - \|\vec q\|^2}}\, 
            \ee^{- \ii c k_0 t}, 
\end{equation}
where $\vec s$ denotes the transverse spatial coordinate in the $xy$-plane and 
\begin{equation}\label{mcAmp}
    \mathcal{A}(\vec k) 
        = \sqrt{\frac{4 \pi k_0}{\rho(\vec k)}}\, 
            \delta \biggl( k_z - \sqrt{k_0^2 - \|\vec q\|^2} \biggr)\, 
            \mathcal{F}(\vec q) 
\end{equation}
in \cref{pwExp}.  The field of \cref{mcMode} has a technical subtlety, however, in that it is not the solution of a differential equation on spacetime, but rather of the \emph{integro}-differential equation 
\begin{equation}\label{wEqPF}
    \biggl( \sqrt{-\vec\grad^2} - \frac{\ii}{c} \pdby{t} \biggr) \Psi^\mc(\vec r, t) = 0, 
\end{equation}
known as the \defn{positive-frequency wave equation}.  But if $\Psi^\mc(\vec r, t)$ belongs to the paraxial regime, so that $\mathcal{F}(\vec q) \approx 0$ unless $\|\vec q\| \ll k_0$, then one can reasonably approximate 
\begin{equation}\label{kzExp}
    \sqrt{k_0^2 - \|\vec q\|^2} = k_0 - \frac{\|\vec q\|^2}{2 k_0} + \cdots 
\end{equation}
in \cref{mcMode}.  Dropping the higher-order terms from this expansion yields the spacetime field \cite{AW, AVNW, AML}.
\begin{equation}\label{paMode}
    \Psi^\pa(\vec s, z, t) 
        = \ee^{\ii k_0 (z - c t)} \int \frac{\ed^2 q}{2 \pi}\, \mathcal{F}(\vec q)\, 
            \ee^{\ii \vec q \cdot \vec s}\, 
            \ee^{- \ii \frac{\|\vec q\|^2}{2 k_0} z}, 
\end{equation}
which is the \defn{paraxial approximation} to \cref{mcMode}.  As noted above, although $\Psi^\pa(\vec s, z, t)$ is still monochromatic, it no longer solves \cref{wEqPF} exactly, nor therefore \cref{wEqIn}, due to the higher-order terms dropped from \cref{kzExp}.

Fields in the paraxial approximation have a number of mathematical features that make them particularly well suited to the description of narrowly collimated, nearly monochromatic beams of classical radiation.  Most importantly, because the arguments of the exponentials in the integrand of \cref{paMode} are all \emph{polynomial} functions of the transverse wave vector $\vec q$, these fields arise from solutions of a (purely) differential equation.  Replacing the constant $k_0$ from \cref{paMode} with $k$ for notational simplicity, we may write these fields in the form 
\begin{subequations}\label{parEnv}
\begin{align}\label{parMod}
    \Psi^\pa(\vec s, z, t) 
        ={}&{} \Xi(\vec s, z)\, \ee^{\ii k (z - c t)} 
\intertext{with}\label{parFour}
    \Xi(\vec s, z) :={}&{} \int \frac{\ed^2 q}{2 \pi}\, \mathcal{F}(\vec q)\, 
            \ee^{\ii \vec q \cdot \vec s}\, 
            \ee^{- \ii \frac{\|\vec q\|^2}{2 k} z} 
\end{align}
\end{subequations}
of a modulated plane wave of \defn{carrier frequency} $ck$.  The envelope function $\Xi(\vec s, z)$ then solves the \defn{paraxial wave equation} 
\begin{align}\label{parEq}
    0 ={}&{} \biggl( 2 \ii k\, \pdby{z} + \triangle \biggr)\, \Xi(\vec s, z) 
        \notag\\[1ex]
        :={}&{} \biggl( 2 \ii k\, \pdby{z} + \pdby[2]{x} + \pdby[2]{y} \biggr) \Xi(\vec s, z)
\end{align}
for that carrier frequency.  Mathematically, \cref{parEq} is the time-dependent Schr\"odinger equation of a (non-relativistic) free particle in two dimensions whose mass is proportional to $k$.  The longitudinal spatial coordinate $z$ plays the role of time in this analogy, and the ``evolution'' of the envelope $\Xi(\vec s, z)$ along the $z$-axis describes the diffractive spreading of the physical beam arising from its localization in the transverse $xy$-plane.

As with any Schr\"odinger-type equation, the ``evolution'' in $z$ produced by \cref{parEq} is unitary in the sense that the inner product 
\begin{align}\label{parIP}
    \eprod[\big]{\Xi_1}{\Xi_2} 
        :={}&{} \int \ed^2 s\, \bar\Xi_1(\vec s, z)\, \Xi_2(\vec s, z) 
        \notag\\
        ={}&{} \int \ed^2 q\, \bar{\mathcal{F}}_1(\vec q)\, \mathcal{F}_2(\vec q)
\end{align}
of solutions $\Xi_{1,2}(\vec s, z)$ with a common carrier frequency is the same for every cross-section (\textit{i.e.}, $z = \mathrm{const.}$) of the beam.  Using this inner product, one can expand the general solution of \cref{parEq} as a superposition of various families of orthonormal modes \cite{MDAB}.  The most common such families used in laser optics \cite{S} involve envelope functions with ``initial data'' at $z = 0$ having the form 
\begin{equation}
    \Xi_\alpha(\vec s, 0) 
        = F_\alpha \biggl( \frac{\vec s}{W} \biggr)\, \frac{\exp - \frac{\|\vec s\|^2}{2 W^2}}{\sqrt{\pi}\, W}, 
\end{equation}
of a Gaussian of width $W$ modulated by one of a \emph{discrete} family of functions $F_\alpha(\frac{\vec s}{W})$ of the transverse coordinates.  Standard examples \cite{PL} of such expansions in laser optics use either the \defn{Hermite--Gauss modes}, where $\alpha = (m, n)$ is a pair of natural numbers and  
\begin{equation}
    F_{mn} \biggl( \frac{\vec s}{W} \biggr) 
        \propto H_m \biggl( \frac{x}{W} \biggr)\, 
            H_n \biggl( \frac{y}{W} \biggr) 
\end{equation}
is a product of Hermite polynomials, or the \defn{Laguerre--Gauss modes}, where $\alpha = (l, p)$ is a pair of integers with $p \ge 0$ and 
\begin{equation}\label{LGmode}
    F_{lp} \biggl( \frac{\vec s}{W} \biggr) 
        \propto \biggl( \frac{s^2}{W^2} \biggr)^{\!|l|/2\,} 
            L^{|l|}_p \biggl( \frac{s^2}{W^2} \biggr)\, 
            \ee^{\ii l \arctan(x, y)}
\end{equation}
is proportional to an associated Laguerre polynomial, with $s := \|\vec s\|$.  The azimuthal phase dependence in the final factor of \cref{LGmode} is often associated with the \emph{orbital} angular momentum of a beam.

The details of these Gaussian mode expansions are not essential in this paper (see \cite{WCZCY} for details).  But we do note that the discrete index $\alpha$ in either case labels different shapes of the phase-front within a modulated beam (see \cite{MDAB}), and that there are well-understood techniques to create, superpose, and filter beams based on this shape information.  These methods, being analogous to those routinely applied to polarization states, open up the possibility of using the wavefront shape to encode quantum information in a wider (\textit{i.e.}, higher-dimensional) channel per photon than the familiar, two-dimensional space of polarization states can offer.  But realizing this advantage in practice hinges on the ability to assign discrete quantum numbers $\alpha$ to \emph{single photons} in quantum (field) theory, rather than to beams of classical radiation.  Moreover, while the inner product of \cref{parIP} gives the space $\mfs{K}_k$ of solutions $\Xi(z, \vec s)$ to the paraxial wave equation (with carrier frequency $ck$) a natural Hilbert space structure, akin to that of the state space of a quantum system, it is important to note that there is nothing inherently quantum mechanical about it\relax
%%%
\footnote{Indeed, if the field $\psi(\vec r, t)$ has units such that the standard action $S[\Psi] := - \frac{1}{2} \int \partial^\mu \Psi\, \partial_\mu \Psi\, \ed^3 x\, \ed t$ for a massless Klein--Gordon field has units of $\hbar$, then the inner product in \cref{parIP} has units of $\hbar c$.  Its value therefore  cannot define a (dimensionless) probability amplitude unless we artificially introduce $\hbar$ into this entirely classical theory.  In contrast, the quantum inner product defined in \cref{qftIP} is dimensionless, and induces the statistical interpretation of the resulting quantum (field) theory.}\relax
%%%
.

\section{Quantum field theory without a fixed basis of modes}\label{S:QFT}

The particle interpretation of the standard quantization of the free, scalar field theory corresponding to \cref{wEqIn} is rooted in its Fock representation \cite{R, GC}.  This representation is built from a single-particle Hilbert space $\mfs{H}$, which is the completion of the space of positive-frequency solutions of the classical wave equation from \cref{pwExp} in the relativistic inner product   
\begin{align}\label{qftIP}
    \iprod{\Psi^+_1}{\Psi^+_2} 
        :={}&{} \frac{\ii}{\hbar c^2} \int \ed^3 r\, \biggl( 
            \bar\Psi^+_1(\vec r, t)\, \pd{\Psi^+_2}{t}(\vec r, t) 
            \notag\\&\hspace{6em}
            - \pd{\bar\Psi^+_1}{t}(\vec r, t)\, \Psi^+_2(\vec r, t) \biggr) 
            \notag\\[1ex]
        ={}&{} \frac{1}{\hbar c} \int \ed^3 k\, \rho(\vec k)\, \bar{\mathcal{A}}_1(\vec k)\, \mathcal{A}_2(\vec k).
\end{align}
This integral is independent of $t$ when $\Psi^+_{1,2}(\vec r, t)$ both solve the wave equation, and becomes positive-definite when restricted to positive-frequency solutions.  The corresponding, multi-particle Fock space $\mfs{F\!H}$ then carries a family of creation operators $\hat a^\dagger \bigl( \Psi^+ \bigr)$ by definition, one for each single-particle state $\ket[\big]{\Psi^+} \in \mfs{H}$, as well as their adjoint annihilation operators $\hat a \bigl( \bar\Psi^+ \bigr) := \bigl[ \hat a^\dagger \bigl( \Psi^+ \bigr) \bigr]{}^\dagger$, one for each adjoint $\bra[\big]{\bar\Psi^+} \in \mfs{H}^*$ of such a single-particle state.  Together, these operators satisfy the canonical commutation relations
\begin{equation}\label{caCCR}
    \comm[\Big]{\hat a \bigl( \bar\Psi^+_1 \bigr)}{\hat a^\dagger \bigl( \Psi^+_2 \bigr)} 
        = \iprod[\big]{\Psi^+_1}{\Psi^+_2}\, \hat 1.
\end{equation}
These relations play the key role of binding the physical interpretation of multi-particle states in the quantum field theory to that of states in the single-particle model.

Textbook presentations of the Fock construction for relativistic fields typically emphasize creation and annihilation operators associated with the elements of a \emph{fixed basis} of single-particle states, rather than with a general element of $\mfs{H}$.  By far the most common choice for this basis consists of the plane-wave momentum eigenstates 
\begin{equation}\label{pwNorm}
    \Phi_{\vec k}(\vec r, t) 
        := \sqrt{\frac{\hbar c}{2 \|\vec k\| \rho(\vec k)}}\, 
            \frac{\ee^{\ii (\vec k \cdot \vec r - c \|\vec k\| t)}}{(2 \pi)^{3/2}}, 
\end{equation}
albeit with varying normalization conventions depending on the preferred choice of the density of states $\rho(\vec k)$.  (See \Cref{fn:Dens} on \cpageref{fn:Dens}.)  These states are orthonormal, in the appropriate sense for any given $\rho(\vec k)$, and the associated creation and annihilation operators therefore satisfy a more familiar form of the canonical commutation relations from \cref{caCCR}: 
\begin{equation}\label{pwOrth}
    \comm[\Big]{\hat a(\vec k_1)}{\hat a^\dagger(\vec k_2)} 
        = \iprod[\big]{\Phi_{\vec k_1}}{\Phi_{\vec k_2}}\, \hat 1
        = \frac{\delta(\vec k_1 - \vec k_2)}{\rho(\vec k_1)}\, \hat 1 
\end{equation}
where $\hat a(\vec k) := \hat a \bigl( \bar\Phi_{\vec k} \bigr)$.  It is important to note, however, that this familiar definition of $\mfs{F\!H}$ is entirely equivalent to the basis-independent version from \cref{caCCR}.

Despite this equivalence, the basis-independent definition of $\mfs{F\!H}$ from \cref{caCCR} has a genuine advantage for our purposes.  Our chief goal is to assess the relative merits of several schemes that have been proposed to associate single-particle quantum states to specific solutions of the paraxial wave equation.  Indeed, while one may well like to fix a basis of solutions to \cref{parEq}, such as the Hermite-- or Laguerre--Gauss modes mentioned above, the different schemes we intend to compare will map that fixed basis to \emph{distinct} sets of single-particle states in $\mfs{H}$.  Fixing a basis in $\mfs{H}$\textit{a priori} would obscure the geometric content of our analysis.  We therefore prefer the basis-independent formulation summarized above.  (See also \cite{DG,D,MS}.)

\section{Paraxial waves as particle modes}

\Cref{S:CPA} has derived the paraxial approximation for scalar fields, leading to the Hilbert space $\mfs{K}_k$ of paraxial wave solutions with a given carrier frequency $ck$, equipped with the inner product from \cref{parIP}.  Meanwhile, \Cref{S:QFT} has reviewed the Fock construction of the corresponding quantum field theory based on the single-particle Hilbert space $\mfs{H}$, equipped with the inner product from \cref{qftIP}.  One can formulate the question of how to associate a single-particle quantum state to a given paraxial wave solution concretely as the search for a suitable mapping $\Xi(\vec s, z) \mapsto \Psi_\Xi(\vec r, t)$ from $\mfs{K}_k$ to $\mfs{H}$.

There are of course many possible mappings of this type.  To be useful physically, however, the mapping we seek should have (at least) the following mathematical properties: 
\begin{enumerate}
\renewcommand\theenumi{\Alph{enumi}}
\setlength\itemsep{1mm}
\item It should be \emph{linear} to ensure that superpositions of single-particle quantum states $\Psi_\Xi(\vec r, t)$ mirror those of the underlying paraxial waves $\Xi(\vec s, z)$.
\item It should be \emph{unitary}, at least in the sense that $\iprod{\Psi_{\Xi_1}}{\Psi_{\Xi_2}}$ vanishes whenever $\eprod{\Xi_1}{\Xi_2}$ does, to ensure that the algebra of projection operators associated with the filtering and measurement of single-particle quantum states $\Psi_\Xi(\vec r, t)$ again mirrors that of the underlying paraxial waves $\Xi(\vec s, z)$.
\item It should be \emph{consistent} in the sense that the constant solution $\Xi_0(\vec s, z) = 1$ of \cref{parEq} is mapped to the carrier wave $\Psi_0(\vec r, t) \propto \ee^{\ii k (z - ct)}$ itself\relax
%%%
\footnote{Note that the constant solution is unique among all solutions of \cref{parEq} in that the spacetime field resulting from \cref{parMod} solves the positive-frequency wave equation \emph{exactly}, and therefore already belongs to $\mfs{H}$.}\relax
%%%
.
\item It should be \emph{covariant} in the sense that rotating $\Xi(\vec s, z)$ about the optical ($+z$-)axis, or rigidly translating it in Euclidean space, induces the same transformation of $\Psi_\Xi(\vec r, t)$ in the Euclidean space of the inertial frame selected by the time coordinate $t$.
\item It should be \emph{scale-invariant} in the sense that the definition of $\Psi_\Xi(\vec r, t)$ in terms of $\Xi(\vec s, z)$ does not introduce any privileged length scale other than that set by the carrier frequency $ck$.
\end{enumerate}
Furthermore, we will restrict our attention to mappings having the general form 
\begin{equation}\label{sbPsi}
    \Psi_\Xi(\vec s, z, t) 
        = \int \frac{\ed^2 q}{2 \pi}\, \mathcal{F}(\vec q)\, 
            \ee^{\ii \vec q \cdot \vec s}\, 
            \ee^{\ii \kappa(\vec q, k) z}\, 
            \ee^{-\ii \omega(\vec q, k) t}, 
\end{equation}
where $\mathcal{F}(\vec q)$ corresponds to $\Xi(\vec s, z)$ via \cref{parFour}, $ck$ is the carrier frequency of $\Xi(\vec s, z)$, and $\kappa(\vec q, k)$ and $\omega(\vec q, k)$ are fixed functions, independent of $\Xi(\vec s, z)$, which remain to be determined.  This class of mappings is broad enough to include all of the candidate mappings that have been proposed in the literature \cite{DG, AW, AVNW, CPB, AML, ACM, SSMs, SSMv}.  These include: 
\begin{enumerate}
\renewcommand\theenumi{\Roman{enumi}}
\setlength\itemsep{1mm}
\item The most obvious mapping from paraxial waves to spacetime fields uses the paraxial approximation itself.  It arises by choosing  
\begin{equation}
    \begin{aligned}
        \kappa^\pa(\vec q, k) &:= k - \frac{\|\vec q\|^2}{2k} 
        \\
        \omega^\pa(\vec q, k) &:= ck
    \end{aligned}
\end{equation}
in \cref{sbPsi}.  While these choices are algebraically simple, however, they do not satisfy \cref{pfdisp}, and therefore do not actually map $\mfs{K}_k$ into the single-particle Hilbert space $\mfs{H}$.  We therefore do not consider this candidate further.
\item A second natural mapping uses the strictly monochromatic fields of \cref{mcAmp}, setting 
\begin{equation}\label{mcdefs}
    \begin{aligned}
        \kappa^\mc(\vec q, k) &:= \sqrt{k^2 - \|\vec q\|^2} 
        \\[1ex]
        \omega^\mc(\vec q, k) &:= ck.
    \end{aligned}
\end{equation}
in \cref{sbPsi}.  The square root here is problematic, however, whenever its argument is negative.  One can certainly impose (the moral equivalent of) a boundary condition, such as by choosing the positive imaginary branch of the root when its argument is negative.  This yields fields that decay exponentially as $z \to +\infty$, corresponding to the physical phenomenon of evanescent fields.  But, since we only consider pure (vacuum) radiation, those same fields necessarily \emph{diverge} exponentially as $z \to -\infty$, and are not normalizable elements of $\mfs{H}$, even in the approximate sense routinely applied to plane waves.  We therefore do not consider this candidate further, either.
\item Aiello and Woerdman \cite{AW, AVNW} have proposed a class of fields corresponding to the choices 
\begin{equation}\label{ipdefs}
    \begin{aligned}
        \kappa^\ip(\vec q, k) &:= k - \frac{\|\vec q\|^2}{2k} 
        \\
        \omega^\ip(\vec q, k) &:= c\, \sqrt{k^2 + \frac{\|\vec q\|^4}{4 k^2}}
    \end{aligned}
\end{equation}
in \cref{sbPsi}.  The resulting spacetime fields $\Psi^\ip_\Xi(\vec r, t)$ have initial data $\Psi^\ip_\Xi(\vec r, 0) = \Psi^\pa_\Xi(\vec r, 0)$, but solve the positive-frequency wave equation exactly throughout spacetime.  We therefore refer to these as \defn{initially paraxial} fields.  Note that the square root in \cref{ipdefs}, in contrast to that in \cref{mcdefs}, can be chosen to be real and positive for all $\vec q$.

\item Sudarshan, Simon and Mukunda \cite{SSMs,SSMv} have proposed the choices 
\begin{equation}\label{hcdefs}
    \begin{aligned}
        \kappa^\hc(\vec q, k) &:= k - \frac{\|\vec q\|^2}{4 k} 
        \\
        \omega^\hc(\vec q, k) &:= c\, \biggl( k + \frac{\|\vec q\|^2}{4 k} \biggr).
    \end{aligned}
\end{equation}
in \cref{sbPsi}.  Adopting their nomenclature, we refer to these as \defn{henochromatic fields}.  These fields have been proposed independently in \cite{DG,AW2}.  Despite the mathematical simplicity associated with the \emph{polynomial} dependence of \emph{both} functions in \cref{hcdefs} on the transverse wave vector $\vec q$, however, their physical content is not so immediately clear.  We will explore this in greater detail below.  
\end{enumerate}
Of course there are infinitely many other choices one could make for the functions $\kappa(\vec q, k)$ and $\omega(\vec q, k)$ in \cref{sbPsi}.  But we will now show that there is in fact a \emph{unique} mapping of that form satisfying all five of the conditions (A through E) laid out above, namely, the mapping to henochromatic fields specified by \cref{hcdefs}.

We begin our uniqueness proof by noting that \emph{any} mapping having the form of \cref{sbPsi} satisfies the linearity condition (A) because $\Psi_\Xi(\vec s, z, t)$ depends linearly on $\mathcal{F}(\vec q)$, which is the Fourier transform of the ``initial data'' $\Xi(\vec s, 0)$ for the ``evolution'' of \cref{parEq}. The Fourier transform is linear, so the mapping $\Xi(\vec s, z) \mapsto \Psi_\Xi(\vec s, z, t)$ is as well.  This implements condition (A).

We defer discussion of the unitarity condition (B) for the moment.

Consider $\mathcal{F}(\vec q) \propto \delta(\vec q)$ in \cref{sbPsi}, the two-dimensional Fourier transform of the ``initial data'' $\Xi_0(\vec s, 0) = 1$.  Demanding that \cref{sbPsi} yields the carrier wave in this case amounts to 
\begin{equation}\label{kwzero}
    \kappa(\vec 0, k) = k
    \quad\text{and}\quad 
    \omega(\vec 0, k) = c k.
\end{equation}
Simply fixing these values at $\vec q = \vec 0$ therefore implements the consistency condition (C).

Now note that, for any fixed $z$ and $t$ in \cref{sbPsi}, $\Psi_\Xi(\vec s, z, t)$ is the two-dimensional (inverse) Fourier transform of a function that is directly proportional to $\mathcal{F}(\vec q)$.  But shifting $\Xi(\vec s, z)$ by some fixed displacement in $\vec s$ multiplies $\mathcal{F}(\vec q)$ by a phase factor, which then induces an identical translation of $\Psi_\Xi(\vec s, z, t)$ in $\vec s$.  It follows that the restriction to mappings of this form automatically guarantees the ``translation part'' of the covariance condition (D).  Meanwhile, the ''rotation part'' of that condition amounts to requiring that there should be no preferred axis in the ($xy$-)plane perpendicular to the optical axis.  That is, $\kappa(\vec q, k)$ and $\omega(\vec q, k)$ should be \emph{isotropic}, depending on $\vec q$ only through its norm $\norm{\vec q}$.  This restriction therefore implements all of the covariance condition (D).

The scale-invariance condition (E) asserts that the only dimensional quantities we can use to define $\kappa(\vec q, k)$ and $\omega(\vec q, k)$ are $\vec q$ and $k$ themselves.  Combined with the previous restriction due to covariance, it follows that $\kappa(\vec q, k)$ and $\omega(\vec q, k)$ should both be (pure-number) functions of the dimension\emph{less} ratio $\norm{\vec q} / k$, apart from overall dimensional factors needed to be compatible with \cref{kwzero}.

Finally, the fields from \cref{sbPsi} must belong to the single-particle Hilbert space $\mfs{H}$ of the quantum theory, meaning that they should solve the positive-frequency wave equation.  They will, provided that 
\begin{equation}\label{pfdisp}
    \omega(\vec q, k) = c \sqrt{\|\vec q\|^2 + \kappa^2(\vec q, k)}.
\end{equation}
We can therefore summarize all of the preceding constraints on $\kappa(\vec q, k)$ and $\omega(\vec q, k)$ by restricting attention to pairs of functions having the forms 
\begin{equation}\label{kweta}
    \begin{aligned}
        \kappa(\vec q, k) 
            &= \|\vec q\| \sinh \eta \biggl( \frac{\|\vec q\|}{k} \biggr) 
            \\[1ex]
        \omega(\vec q, k) 
            &= c \|\vec q\| \cosh \eta \biggl( \frac{\|\vec q\|}{k} \biggr), 
    \end{aligned}
\end{equation}
where $\eta(r)$ is a pure-number valued function, yet to be determined, of a pure number $r \ge 0$.  Note that $\eta(r)$ must diverge logarithmically as $r \to 0$ for \cref{kweta} to approach \cref{kwzero}.

Our sole remaining task is to implement the unitarity condition (B).  To that end, let $\Xi_{1,2}(\vec s, z)$ denote solutions of the paraxial wave equation with carrier frequencies $ck_{1,2}$, respectively.  Define the solutions $\Psi_{1,2}(\vec s, z, t)$ of the positive-frequency wave equation using \cref{sbPsi} for some pair of functions $\kappa(\vec q, k)$ and $\omega(\vec q, k)$ having the form of \cref{kweta}.  We can then simply compute the inner product from \cref{qftIP} for these single-particle quantum states:
\begin{widetext}
\begin{align}\label{sbIP}
    \iprod{\Psi_1}{\Psi_2} 
        &= \frac{2 \pi}{\hbar c^2} \int \ed^2 q\, 
            \bar{\mathcal{F}}_1(\vec q)\, 
            \mathcal{F}_2(\vec q)\, 
            \delta \bigl( \kappa(\vec q, k_2) - \kappa(\vec q, k_1) \bigr)\, 
            \ee^{\ii [\omega(\vec q, k_1) - \omega(\vec q, k_2)] t}
            \bigl( \omega(\vec q, k_2) + \omega(\vec q, k_1) \bigr)
            \notag\\[1ex]
        &= \frac{4 \pi}{\hbar c^2}\, \delta (k_1 - k_2) \int \ed^2 q\, 
            \bar{\mathcal{F}}_1(\vec q)\, \mathcal{F}_2(\vec q)\, 
            \frac{\omega(\vec q, k_1)}{\bigl| \pd{\kappa}{k}(\vec q, k_1) \bigr|}.
\end{align}
\end{widetext}
It follows that $\Psi_{\Xi_{1,2}}(\vec s, z, t)$ are orthogonal whenever the underlying paraxial waves have the different carrier frequencies $ck_1 \ne ck_2$.  Furthermore, when they do have the same carrier frequency, the remaining integral in \cref{sbIP} is proportional to the paraxial approximation's inner product integral from \cref{parIP} if and only if 
\begin{equation}\label{kwucon}
    \frac{\omega(\vec q, k)}{\bigl| \pd{\kappa}{k}(\vec q, k) \bigr|} 
        = \frac{ck}{\Bigl| \frac{\|\vec q\|}{k}\, \eta' \bigl( \frac{\|\vec q\|}{k} \bigr) \Bigr|} 
        = \Omega(k) .
\end{equation}
for some function $\Omega(k)$, yet to be determined, which must be independent of $\vec q$.  The second expression here uses the forms of $\kappa(\vec q, k)$ and $\omega(\vec q, k)$ from \cref{kweta}.  Now, the left side of \cref{kwucon} depends on $\|\vec q\|$ \emph{only} through the ratio $\|\vec q\| / k$, whereas its right side depends only on $k$.  It follows that the left side must be independent of $\|\vec q\|$, and therefore also of $k$, and thus the right side must be independent of $k$ as well.  In other words, there must be a constant $\beta$ such that 
\begin{equation}
    \eta'(r) = - \frac{\beta}{r} 
    \quad\text{and}\quad
    \Omega(k) = \frac{ck}{|\beta|}.
\end{equation}
Solving the resulting, first-order ordinary differential equation for $\eta(r)$ yields 
\begin{equation}\label{kwunit}
    \begin{aligned}
        \kappa(\vec q, k) 
            &= \frac{k^\beta\, \|\vec q\|^{1 - \beta}}{2 \ee^{-\alpha}} 
                - \frac{\|\vec q\|^{1 + \beta}}{2 \ee^\alpha\, k^\beta} 
        \\[1ex]
        \frac{\omega(\vec q, k)}{c} 
            &= \frac{k^\beta\, \|\vec q\|^{1 - \beta}}{2 \ee^{-\alpha}} 
                + \frac{\|\vec q\|^{1 + \beta}}{2 \ee^\alpha\, k^\beta} 
    \end{aligned}
\end{equation}
upon substitution into \cref{kweta}, where $\alpha$ is a constant of integration.  \Cref{kwzero} then shows that $\beta = +1$ and $\alpha = \ln 2$.  We therefore conclude, as claimed previously, that the \emph{only} choice of $\kappa(\vec q, k)$ and $\omega(\vec q, k)$ in \cref{sbPsi} that leads to a mapping $\Xi(\vec s, z) \mapsto \Psi_\Xi(\vec s, z, t)$ satisfying the conditions (A--E) laid out above is that of \cref{hcdefs}.

\section{Completeness of henochromatic states}

The previous section has shown that the mapping from solutions of the paraxial wave equation to henochromatic single-particle quantum states satisfies 
\begin{equation}
    \iprod[\big]{\Psi^\hc_1}{\Psi^\hc_2} 
        = \frac{4 \pi k_1}{\hbar c}\, \delta(k_1 - k_2)\, \eprod[\big]{\Xi_1}{\Xi_2}.
\end{equation}
In particular, orthogonal paraxial waves $\Xi_{1,2}(\vec s, z)$, or arbitrary paraxial waves having distinct carrier frequencies $c k_1 \ne c k_2$, give rise to orthogonal quantum states.  In this section, we show that the set of all henochromatic quantum states arising in this way is also \emph{complete} in the single-particle Hilbert space $\mfs{H}$.  That is, \emph{any} single-particle quantum state can be written (uniquely) as a superposition of henochromatic states, each of which derives, as described above, from a specific solution of the paraxial wave equation with a specific carrier frequency.  This may seem surprising at first since the paraxial approximation is only expected to be useful physically for modeling a subset of radiation fields: those lying in the paraxial regime.  But in fact this result is simply and directly connected to natural geometric structures on spacetime.

Following \cite{SSMs}, consider the spacetime coordinate transformation that replaces the inertial coordinates $t$ and $z$ with the null coordinates 
\begin{equation}
    u := z - ct 
    \qquad\text{and}\qquad
    v := \tfrac{1}{2} \bigl( z + ct \bigr), 
\end{equation}
with the $x$ and $y$ coordinates unchanged.  \Cref{wEqIn} takes the form 
\begin{equation}
    \Box \Psi(\vec s, v, u) 
      = \biggl( \pdby[2]{x} + \pdby[2]{y} + 2\, \pdby[2][1]{u}{v} \biggr)\, \Psi(\vec s, v, u) = 0
\end{equation}
in these null coordinates.  Separating variables by expanding $\Psi(\vec s, v, u)$ as a Fourier transform in $u$,  
\begin{subequations}\label{uexp}
\begin{align}\label{umod}
    \Psi(\vec s, v, u) = \int \ed k\, \ee^{\ii k u}\, \Xi(\vec s, v; k) 
\intertext{produces the reduced wave equation}\label{upareq}
    \biggl( 2 \ii k\, \pdby{v} + \triangle \biggr)\, \Xi(\vec s, v; k) = 0.
\end{align}
\end{subequations}
This is precisely the paraxial wave equation from \cref{parIP}, albeit with the role of the longitudinal coordinate $z$ on space now played by the null coordinate $v$ on spacetime.  Expanding $\Xi(\vec s, v; k)$ in a 2-dimensional Fourier transform, as in \cref{parFour}, shows that a general solution of the wave equation can be written in the form 
\begin{align}\label{uvPsi}
    \Psi(\vec s, v, u) 
        &= \int \frac{\ed k\, \ed^2 q}{2 \pi}\, \mathcal{F}(\vec q; k)\, 
            \ee^{\ii \vec q \cdot \vec s}\, 
            \ee^{\ii k u}\, 
            \ee^{- \ii \frac{\|\vec q\|^2}{2 k} v}
            \\[1ex]\notag
        &= \int \frac{\ed k\, \ed^2 q}{2 \pi}\, \mathcal{F}(\vec q; k)\, 
            \ee^{\ii \vec q \cdot \vec s}\, 
            \ee^{\ii \kappa^\hc(\vec q, k) z}\, 
            \ee^{- \ii \omega^\hc(\vec q, k) t}.
\end{align}
Note that $\omega^\hc(\vec q, k) > 0$ in \cref{hcdefs} is positive if and only if $k > 0$.  It follows that restricting the integral here to positive carrier frequencies is equivalent to restricting to the positive-frequency fields of \cref{pwExp}, which span the single-particle quantum Hilbert space. 
 Thus, every single-particle quantum state can be written (uniquely) as a superposition of henochromatic fields.

\section{Discussion}

Our principal result in this paper is that the mapping from solutions $\Xi(\vec s, z)$ of the paraxial wave equation to henochromatic single-particle states $\Psi_\Xi^\hc(\vec s, z, t)$ on spacetime is \emph{unique} within a large class of similar mappings in that it preserves both superpositions and projections of the states of a paraxial beam.  Physically, this implies that the logic of both the preparation and the subsequent measurement of henochromatic quantum states exactly mirrors standard resolutions of a classical beam in the paraxial approximation into orthogonal modes corresponding to different shapes of the wavefront within the beam.  We have also reviewed how the paraxial wave equation emerges naturally, without any approximation, by recasting the relativistic wave equation in appropriate null coordinates \cite{Drd}.  It follows that any choice of an orthogonal basis of modes for the paraxial wave equation (for each carrier frequency $ck$) naturally gives rise to a \emph{complete} basis of single-particle quantum states.  While the natural, mathematical connection between henochromatic fields and the paraxial wave equation has been emphasized before at the level of the field equation \cite{SSMs, SSMv}, we believe that the above analysis of their quantum mechanical unitarity and completeness is new.

We close with a comment.  Despite their advantages highlighted above, henochromatic fields at first seem to have a disadvantage as well in that their structures as spacetime fields do not necessarily resemble the corresponding fields in the paraxial approximation.  Indeed, there are well-understood experimental methods to construct non-trivial paraxial waves, such as the higher-order Hermite-- or Laguerre--Gauss modes, by passing a laser beam through certain optical elements (structured diffraction gratings, spiral wave plates, \textit{etc.}) \cite{S}.  But these constructions are rooted mathematically in the paraxial approximation.  How should one adapt these methods to produce the \emph{henochromatic} states corresponding to those non-trivial paraxial waves instead?  We now argue that in fact no modification is necessary.

Compare the spacetime fields 
\begin{subequations}\label{fldcomp}
\begin{align}\label{fldpa}
    \Psi_\Xi^\pa(\vec s, z, t) &:= \Xi(\vec s, z)\, \ee^{\ii k (z - ct)} 
\intertext{and}\label{fldhc}
    \Psi_\Xi^\hc(\vec s, z, t) &:= \Xi \bigl( \vec s, \tfrac{1}{2} (z + c t) \bigr)\, \ee^{\ii k (z - ct)} 
\end{align}
\end{subequations}
arising in the paraxial approximation and the henochromatic approach, respectively, from a common solution $\Xi(\vec s, z)$ of the paraxial wave equation with carrier frequency $ck$.  While replacing $z \mapsto \frac{1}{2} (x + ct)$ in the latter expression implies that these fields differ at most points of spacetime, note that they do resemble one another in the region $z \approx ct$.  Moreover, if $\Xi(\vec s, z)$ is deep in the paraxial regime in the sense that its Fourier transform $\mathcal{F}(\vec q)$ at $z = 0$ has the bulk of its support in the region $\|\vec q\| \ll k$, then $\Xi(\vec s, z)$ itself varies far more slowly in the longitudinal direction than in the transverse directions.  It follows that the longitudinal extent of the region around $z = ct$ where the two fields of \cref{fldcomp} approximate one another can be quite large in precisely those instances where the paraxial approximation is most accurate.

To make the preceding point more precise, take $\mathcal{F}(\vec q; k)$ in \cref{uvPsi} not only to be narrowly peaked around $\vec q = \vec 0$, but to be even more narrowly peaked around $k = k_0$ in the carrier frequency.  If $\mathcal{F}(\vec q; k)$ varies like a Gaussian in $k$, for example, then the spacetime field of \cref{uvPsi} describes a longitudinal pulse propagating along the optical axis.  The longitudinal extent of the pulse is large compared to its cross-sectional width, which in turn is large compared to the wavelength of the radiation within.  But within its slowly varying, longitudinal, Gaussian envelope, the pulse field will be well approximated by the field $\Psi_{\Xi_0}^\pa(\vec s, z, t)$ arising in the paraxial approximation, where $\Xi_0(\vec s, z)$ is the solution of the paraxial wave equation corresponding to $\mathcal{F}_0(\vec q) := \mathcal{F}(\vec q; k_0)$.  That is, wherever the amplitude of such a pulse field is non-negligible, it resembles a field in the paraxial approximation.  In particular, any optical element designed to act on a Gaussian beam in the paraxial approximation to yield higher-order paraxial beam modes should continue to operate on such pulses constructed from henochromatic fields in essentially the same way.  Henochromatic pulses can therefore resemble fields in the paraxial approximation, not everywhere in spacetime, but \emph{everywhere that matters}.  At the same time, they solve the (positive-frequency) wave equation exactly, and therefore define proper, normalizable, single-particle states in the quantum theory.  In this sense, they provide a natural bridge from the classical optics of paraxial beams to the quantum optics of single photons.

\section* {Acknowledgments}

CB would like to thank Minerba Betancourt, Gerardo Estrada, Angela Guzman, Grigoriy Kreymerman, Vasudevan Lakshminarayanan, Warner Miller, Bill Rhodes and Ayman Sweiti for discussions of and suggestions regarding this work.


\begin{thebibliography}{MM}

\bibitem{SSMs}
E.C.G.~Sudarshan, R.~Simon and N.~Mukunda.
Paraxial-wave optics and relativistic front description.  I.  The scalar theory.
\textit{Phys.\ Rev.\ A} \textbf{28}, 2921--2932 (1983).

\bibitem{SSMv}
E.C.G.~Sudarshan, R.~Simon and N.~Mukunda.
Paraxial-wave optics and relativistic front description. II. The vector theory.
\textit{Phys.\ Rev.\ A} \textbf{28}, 2921--2932 (1983).

\bibitem{CPB}
G.F.~Calvo, A.~Pic\'on and E.~Bagan.
Quantum field theory of photons with orbital angular momentum.
\textit{Phys.\ Rev.\ A} \textbf{73}, 013805 (2006).

\bibitem{ACM}
Arvind, S.~Chaturvedi and N.~Mukunda.
On ‘orbital’ and ‘spin’ angular momentum of light in classical and quantum theories --- a general framework.
\textit{Fortschr.\ Phys.}\ \textbf{66}, 1800040 (2018).

\bibitem{AW}
A.~Aiello and J.P.~Woerdman.
Exact quantization of a paraxial electromagnetic field.
\textit{Phys.\ Rev.\ A} \textbf{72}, 060101(R) (2005).

\bibitem{AW2}
A.~Aiello and J.P.~Woerdman.
Exact Paraxial Quantization II: Developments.
\texttt{arXiv:quant-ph/0509082}

\bibitem{AML}
A.~Aiello, C.~Marquardt and G.~Leuchs. 
Transverse angular momentum of photons. \textit{Phys.\ Rev. \ A} {\bf 81}, 053838 (2010).

\bibitem{AVNW}
A.~Aiello, J.~Visser, G.~Nienhuis and J. P.~Woerdman.
Angular spectrum of quantized light beams.
\textit{Opt.\ Lett.}\ \textbf{31} (2006) 525.

\bibitem{DG}
I.H.~Deutsch and J.C.~Garrison.
Paraxial quantum propagation.
\textit{Phys.\ Rev.\ A} \textbf{43}, 2498--2513 (1991).

\bibitem{Vectorpaper}
M.F.\ \surname{Jongewaard de Boer} and C. Beetle.
On the paraxial approximation in quantum optics II: Henochromatic modes of a Maxwell field.
Submitted to \textit{Phys.\ Rev.\ A}.

\bibitem{MB}
E.W.~Max Born, 
\textit{Principles of optics}. Cambridge Uni-
versity Press, (1997).

\bibitem{EW}
E.~Wolf, 
\textit{Introduction to the Theory of Coherence and Polarization of Light}. Cambridge University Press, (2007).

\bibitem{MDAB}
D.A.B.~Miller. 
Waves, modes, communications, and optics: a tutorial. \textit{Adv.\ in \ Opt.\ and \ Phot.} Vol. 11, Issue 3, Pages 679-825 (2019).

\bibitem{S}
A.E.~Siegman.
Lasers.  (University Science Books, Mill Valley, California, 1986).

\bibitem{PL}
V.~Pe\v{r}inov\'{a} and A.~Luk\v{s}. 
Quantization of Hermite--Gaussian and Laguerre--Gaussian beams and their spatial transformations.
\textit{J.\ of \ Mod.\ Opt.}\ {\bf 53}, 659–-675 (2006).

\bibitem{WCZCY}
Y.~Wang, Y.~Chen, Y.~Zhang, H.~Chen and S.~Yu. 
Generalised Hermite–Gaussian beams and mode transformations. \textit{J.\ of \ Opt.} {\bf 18}, 055001 (2016).

\bibitem{R}
L.H.~Ryder.
\textit{Quantum Field Theory}, Cambridge University Press, (1996).

\bibitem{GC}
J. C.~ Garrison and R. Y.~ Chiao, 
\textit{Quantum Optics}. Oxford University Press, (2008).

\bibitem{D}
I. H.~Deutsch. 
A basis‐independent approach to quantum optics. \textit{Am.\ J.\ of \ Phys.} {\bf 59}, 834 (1991).

\bibitem{MS}
N.~Mukunda and E.C.G.~Sudarshan.
The three faces of Maxwell's equations.
\textit{Pram\={a}na -- J.\ Phys.}\ \textbf{27}, 1--18 (1986).

\bibitem{Drd}
P.A.M.~Dirac.
Forms of Relativistic Dynamics.
\textit{Rev.\ Mod.\ Phys.} \textbf{21}, 392--399 (1949).




%\bibitem{BBMM}
%C.~Beetle, M.~Betancourt, J.R.~McDonald and W.A.~Miller.
%A Maximal Bound on the Efficiency of OAM Beam Sorting.
%To appear.
%
%\bibitem{BBv}
%C.~Beetle and M.~Betancourt.
%Exploiting the Paraxial Approximation in Quantum Optics: Quantization of Scalar And Vector Fields in Henochromatic Modes.
%To appear.

% \bibitem{SW}
% R.F.~Streater and A.S.~Wightman.
% \textit{PCT, Spin and Statistics, and All That.}
% New York: W.A.~Benjamin, 1964.

% \bibitem{A20}
% Aiello, A., 2020. Field theory of monochromatic optical beams: I. Classical fields. Journal of Optics 22, 014001.. doi:10.1088/2040-8986/ab5c5c

% \bibitem{W}
% R.M.~Wald.
% \textit{Quantum Field Theory in Curved Spacetime and Black Hole Thermodynamics.}
% Chicago: University of Chicago Press, 1994.

% \bibitem{U}
% W.G.~Unruh.
% Notes on black-hole evaporation.
% \textit{Phys.\ Rev.\ D} \textbf{14} (1976) 870--892.

% \bibitem{H}
% S.W.~Hawking.
% Particle creation by black holes.
% \textit{Commun.\ Math.\ Phys.}\ \textbf{43} (1975) 199--220.





\end{thebibliography}
\end{document}